\documentclass{aa}  

\usepackage{graphicx}
\usepackage{txfonts}
\usepackage{lipsum}
\usepackage{amsmath}
\usepackage{lscape}             
\usepackage{placeins}           
\usepackage{natbib}
\usepackage{xcolor}
\usepackage{hyperref} 
\hypersetup{
    colorlinks=true,
    linkcolor=blue,
    citecolor=blue
    }

\newcommand{\hrieuv}{HRI\textsubscript{EUV}\xspace}

\def\arcsec{$^{\prime\prime}$}
                                
\begin{document}

   \title{Evidence for Suppressed Non-thermal Broadening in a Flare Ribbon from Stereoscopic Observations}

%
%
%

   \author{Ziwen Huang\inst{1}\corrauth{huangz@mps.mpg.de}        
        \and 
        L. P. Chitta\inst{\ref{i:mps}}
        \and L. Teriaca\inst{\ref{i:mps}}
        }

   \institute{Max Planck Institute for Solar System Research, Justus-von-Liebig-Weg 3, 37077 G\"ottingen, Germany\label{i:mps}}

   \date{Received MM DD, 2026}

 
  \abstract
   {Flare ribbons are footpoints of newly reconnected magnetic loops. They are commonly associated with strong plasma flows and pronounced non-thermal broadening of transition-region lines, resulting from impulsive energy deposition. However, the mechanism governing energy transport into flare ribbons remains uncertain, particularly under different magnetic field configurations.}
   {We aim to understand the energy deposition process at a flare ribbon located close to a sunspot by determining the physical nature of the small‑scale downflows hosted by the ribbon.} 
   {We study small-scale downflows in a remote ribbon during the M7.7-class event on 30 September 2024, using coordinated observations from the Solar Dynamics Observatory, the Interface Region Imaging Spectrograph (IRIS), and Solar Orbiter. We use stereoscopic imaging from Solar Orbiter and IRIS to detect plane-of-sky motions on small scales ($\sim$200\,km). We conduct spectroscopic analysis of IRIS Si~{\sc iv} profiles, complemented by C~{\sc ii} and Mg~{\sc ii} diagnostics.}
  {Apparent downflow speeds measured from Solar Orbiter imaging (limb view) are 85--130\,km\,s\,$^{-1}$, while IRIS images (disk view) reveal plane-of-sky speeds of 70--85\,km\,s$^{-1}$. In contrast, the IRIS Si~{\sc iv} profiles at the brightest ribbon pixels are well approximated by a single Gaussian, with minimal Doppler shifts and unexpectedly small non‑thermal widths ($\sim$10\,km\,s\,$^{-1}$). Peak Si~{\sc iv} emission also coincides with deeper Mg~{\sc ii} central reversal, while the electron densities are one to two orders of magnitude higher than typical quiet sun values.}
  {The absence of significant line-of-sight velocity and non-thermal broadening differs from the typical behavior of flare ribbons. Our results are consistent with a scenario in which plasma motions are guided by highly inclined (nearly horizontal) magnetic fields in chromospheric canopy and are not accompanied by strong turbulence or wave motions. This study suggests that flare ribbons can exhibit a wide diversity of lower-atmospheric responses and emphasizes the critical role of magnetic field geometry.}

   \keywords{Sun: chromosphere -- Sun: flares -- Sun: transition region -- Sun: UV radiation}

   \maketitle
   \nolinenumbers

\section{Introduction}
\label{introduction}

Solar flares are among the most energetic phenomena in the solar atmosphere, releasing magnetic energy through reconnection and converting it into heating, bulk plasma motions, and particle acceleration. A significant fraction of this energy is transported from the reconnection site to the chromosphere via thermal conduction or non-thermal particles, heating the chromosphere and producing bright flare ribbons observed in H$\alpha$ and ultraviolet (UV) radiation. These ribbons mark the footpoints of the evolving post-flare magnetic field \citep[e.g.,][]{2012A&A...543A.110A, 2013A&A...555A..77J, 2016A&A...591A.141J} and are associated with hard X‑ray (HXR) footpoints caused by collisions of accelerated electrons with dense chromospheric plasma \citep{2003ApJ...595L.107E, 2007ApJ...654..665T}. 

The connection between flare ribbons and energy release in the lower atmosphere has been established through spectroscopic observations, e.g., of Doppler shifts. When energy is deposited in the chromosphere, the heated plasma expands upward, producing blueshifts of spectral lines, and fills the post-flare loops. This chromospheric ablation (or evaporation) \citep{1968ApJ...153L..59N, 1982ApJ...263..409A}, generates upflows that can reach hundreds of km\,s$^{-1}$ in hot coronal lines \citep[e.g.,][]{1982MmSAI..53..495A, 1985SoPh...96..129A, 2004ApJ...613..580B, 2006SoPh..234...95D, 2015ApJ...811....7L}. Below the hotter upflow region, cooler plasma is  driven downward, producing redshifts or red wing enhancements in lines formed in the chromosphere and transition region. This is termed chromospheric condensation \citep{1985ApJ...289..414F}. Downflows originate from the same location as upflows, with speeds typically an order of magnitude smaller than that of upflows, ranging from a few to a few tens of km\,s$^{-1}$ \citep[e.g.,][]{1994ApJ...424..459W, 2003ApJ...588..596T, 2006A&A...455.1123T, 2022FrASS...940945L}.

In flare ribbons, spectral lines also exhibit broadening beyond what can be explained by thermal and instrumental effects. This non-thermal broadening, especially in optically thin plasma, has been reported as a consequence of turbulence \citep{1977ApJ...212L.143D, 2020A&A...639A..21P}, acoustic or magnetohydrodynamics (MHD) waves \citep{1978ApJ...226..698M, 1991SoPh..131...41M}, and the superposition of different unresolved velocity components \citep{2011ApJ...740...70M, 2015ApJ...803...84P}. Significant non-thermal broadening, corresponding to speeds of about 100\,km\,s$^{-1}$ (full width at half maximum (FWHM) of the spectral lines), has been reported in hot lines (>\,10\,MK) such as Fe~{\sc xxiii} and Fe~{\sc xxiv} \citep{2011ApJ...740...70M, 2013ApJ...766..127Y}. A positive correlation between blueshift and non‑thermal widths was reported in the Fe~{\sc xv} and Fe~{\sc xvi} (2\,MK) lines at the loop footpoint \citep{2011ApJ...740...70M}, as well as in the case of Fe~{\sc xxi} line at 10\,MK \citep{2015ApJ...803...84P}. This broadening is often attributed to chromospheric evaporation. However, \citet{2019ApJ...879L..17P} conclude that the superposition of unresolved evaporative flows is inconsistent with observations of very hot flare plasma ($\sim$11~MK), as their models could not reproduce the symmetric Fe~{\sc xxi} profiles observed by Interface Region Imaging Spectrograph \citep[IRIS;][]{2014SoPh..289.2733D}. The scenario in the transition region and chromosphere may be different. For example, line profiles of optically thick plasma are more complex as their broadening is also contributed by opacity effect. 

Transition region plasma traced by the Si~{\sc iv} line observed by IRIS also shows line broadening associated with flares \citep{2015ApJ...811..139T, 2023ApJS..268...62W}. \cite{2023ApJS..268...62W} also report central reversal in the Si~{\sc iv} line during an X‑class flare, which may indicate that Si~{\sc iv} line can become optically thick due to non-thermal electrons from a strong flare. \cite{2018SciA....4.2794J} show that in a small B-class flare ribbon, the non‑thermal width of Si~{\sc iv} begins to increase (from 9 to 30\,km\,s$^{-1}$, FWHM) before the line intensity rises. This early onset suggests that turbulence might contribute to the lower atmosphere heating. While the line widths are relatively modest in this case, substantially broader profiles have been observed in other flare ribbons, with non-thermal widths exceeding 100\,km\,s$^{-1}$ \citep{2020ApJ...890L...2C}. With high‑cadence observations, \cite{2022FrASS...940945L} further identify that the non‑thermal broadening of Si~{\sc iv} is correlated with a secondary, redshifted component (up to 70\,km\,s$^{-1}$), which they interpret as downflowing condensation. 

The structure of flare ribbons is not uniform. For example, H$\alpha$ kernels, appearing as localized bright patches within ribbons, are considered the sites most directly connected to energy release \citep{1986Ap&SS.118..149K, 1993A&A...271..292D, 2002ApJ...578L..91A, 2014ApJ...788L..18S}. Similarly, high‑resolution Extreme Ultraviolet (EUV) observations from the Solar Dynamics Observatory \citep[SDO;][]{2012SoPh..275....3P} reveal that ribbons are composed of bright knots slipping along the ribbon \citep{2016ApJ...823...41D}. Using IRIS data, \cite{2015ApJ...807L..22G} find that these flare kernels are associated with coronal upflows and chromospheric downflows, i.e., the chromospheric evaporation and condensation discussed above. Moreover, these fine structures are thought to originate from the current sheet tearing \citep{2021ApJ...920..102W, 2025AAS...24622404J} and may be associated with the presence of the non-thermal electrons that produce HXR emission \citep{2002ApJ...578L..91A, 2022FrASS...940945L, 2023ApJ...954....7L}. 

Although flare ribbons, especially the fine structures within, clearly indicate impulsive energy deposition in the lower atmosphere, how different magnetic environments influence the imaging and spectral signatures remains unclear. A valuable opportunity to address this question arises from an M7.7-class flare that occurred on 30 September 2024, observed near the west limb by the High Resolution Imager of the Extreme Ultraviolet Imager \citep[\hrieuv;][]{2020A&A...642A...8R} onboard Solar Orbiter \citep{2020A&A...642A...1M}, when the spacecraft was close to its perihelion. This event was investigated in detail by \cite{2026A&A...705A.113C}, who suggest that the flare developed through an avalanche-like self-organized process. Their interpretation is supported by the presence of rapid, small-scale reconnection events occurring on timescales of seconds prior to the main energy release. Owing to the complex magnetic configuration of the event, the flare exhibited multiple ribbons. The authors further report ribbon-like emission patterns formed by raining plasma blobs, suggesting energy transport along magnetic field lines toward the lower atmosphere.

One of the flare ribbons of the above flare is well suited for diagnosing energy deposition as part of the flare ribbon is covered by IRIS, which provides spectral coverage of lines formed from chromospheric to transition region temperatures. These IRIS data are complemented by high‑resolution \hrieuv observations and multiple EUV channels from the Atmospheric Imaging Assembly \citep[AIA;][]{2012SoPh..275...17L} on board SDO, giving us a comprehensive multi‑wavelength view. In addition, the viewing angles of Solar Orbiter and Earth differed by approximately 107\textdegree, allowing stereoscopic observations of the same flare structures in quadrature. Moreover, because this particular ribbon lies close to the penumbra of a nearby sunspot, the dataset also lets us examine how flare ribbon energy deposition varies under different magnetic environments.

This paper is structured as follows. A detailed overview of the observations can be found in Section~\ref{overview}. Section~\ref{downflow} focuses on the small-scale downflows and their plane-of-sky (POS) motion measurements. Section~\ref{Spectral} presents the analysis of the IRIS spectra, followed by the discussions on how to reconcile all the signatures in Section~\ref{Discussions} and conclusions in Section~\ref{Conclusions}.

\section{Observation overview}
\label{overview}

\begin{figure*}[htbp]
\centering 
\includegraphics[width=1.0\textwidth]{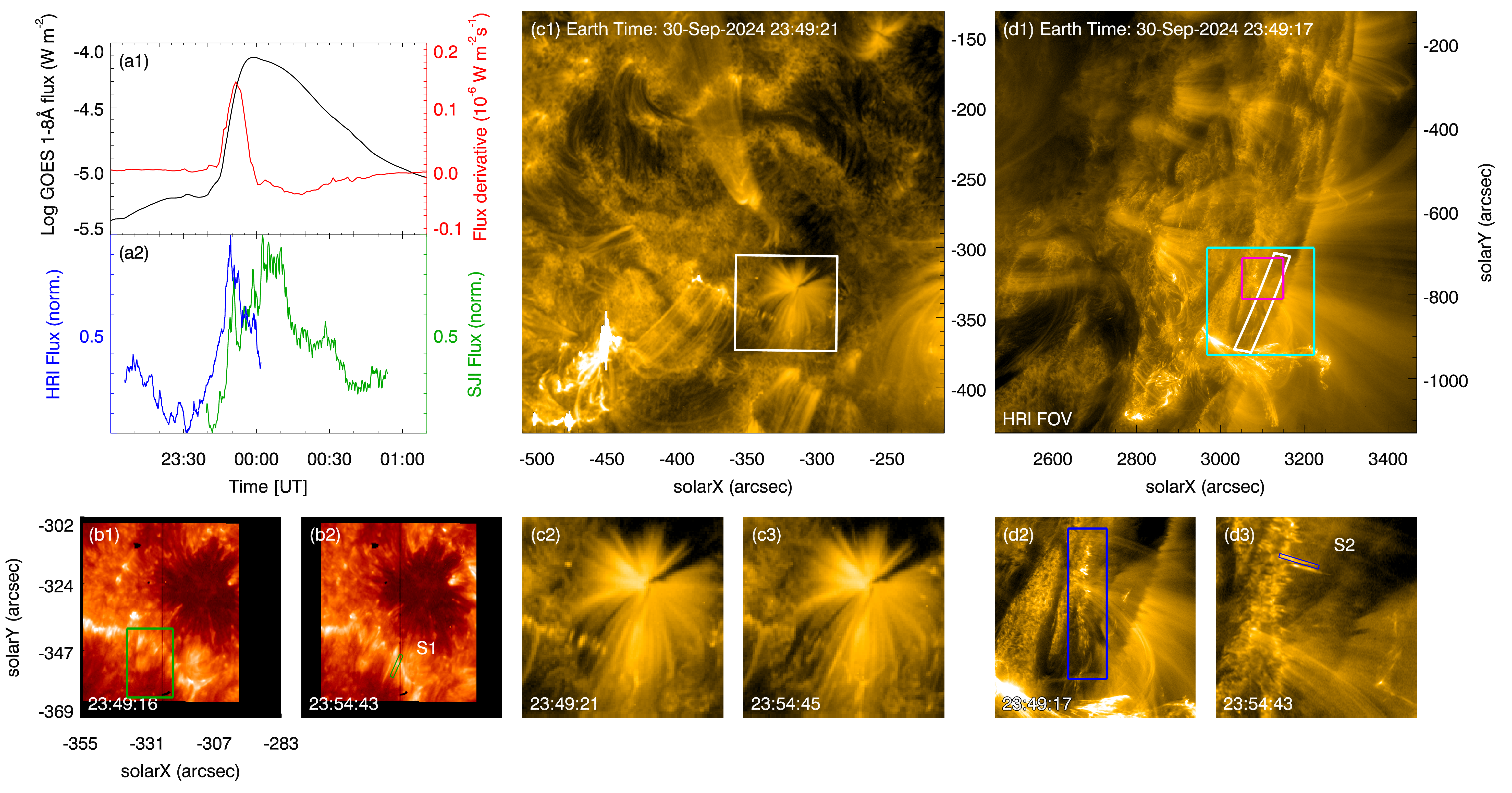}
\caption{Stereoscopic overview of the flare ribbon. (a1) Light curve of the GOES 1--8 \AA\ flux (black), with the flare peaking at around 23:59 UT. The red curve shows the time derivative of the flux. (a2) Light curves of the targeted flare ribbon locations in SJI (green) and \hrieuv (blue). The regions used for these calculations are indicated in (b1) and (d2). (b1)–(b2) Example SJI images (1330~\AA) during the impulsive phase (23:49 UT)  and at a later time (23:54 UT) when the ribbon structure is clear. (c1) AIA 171\AA\ image showing the flare ribbon and its connection to the main flare. The white box marks the SJI FOV. Zoomed-in views corresponding to the times of (b1) and (b2) are shown in (c2) and (c3), respectively. (d1) Full FOV of an \hrieuv image, with the SJI FOV projected and outlined in white. (d2) and (d3) show Zoomed-in views of the cyan and magenta boxes in (d1), respectively, at the same times as (b1) and (b2). Slit S1 in (b2) and slit S2 in (d3) are used to generate the time–slice plots.}
\label{fig1}
\end{figure*}

This flare occurred between 30 September 2024 23:37~UT and 1 October 2024 00:23~UT, peaking at around 23:59~UT (Earth time). The soft X-ray light curve from GOES 1--8 \AA\ shown in Fig.~\ref{fig1}~(a1) together with its derivative, which serves as a proxy for the flare energy release rate, show that the flare peaked between 23:45~UT and 23:59~UT. This event was observed by SDO close to the solar disk center. Fig.~\ref{fig1}~(c1) shows a sub-region of AIA 171 \AA, covering the flaring event in the lower-left corner. AIA 171 \AA\ has a spatial resolution of about 1.5\arcsec\ (with a pixel size of 0.6\arcsec, about 430 km) and a time cadence of 12\,s. Our target ribbon is remotely located north-west of the main flare, connected with bright loops and partly captured by IRIS. The white box in Fig.~\ref{fig1}~(c1) indicates the field of view (FOV) of Slit-jaw Imager (SJI) aboard IRIS. 

In this dataset, IRIS operated in an 8-step coarse raster mode, accompanied by slit-jaw imaging in the 1330 \AA\ channel, which is dominated by C~{\sc ii} emission and thus primarily traces upper chromospheric to lower transition-region plasma. The resolution of the IRIS SJI is about 0.33\arcsec which matches the slit width, corresponding to about 240\,km on the Sun surface. The observations were carried out in low telemetry mode, with an exposure time of 8\,s, raster step cadence of 9.3\,s and raster step size of 2\arcsec. The raster sequence covers multiple spectral lines formed over a broad temperature range, including C~{\sc ii} 1336~\AA, Fe~{\sc xii} 1349~\AA, O~{\sc i} 1356~\AA, Si~{\sc iv} 1403~\AA\ and Mg~{\sc ii}~2796 \AA.

The IRIS slit scanned the target region from east to west. In the western four slit positions, the scanning FOV covers the sunspot and an associated jet-like event located to the north of the sunspot. To avoid contamination from this dynamic jet activity, these four slit positions are excluded from this analysis. The remaining four slit positions capture one end of the remote ribbon, which locates very close to the sunspot penumbra. Fig.~\ref{fig1}~(b1) and (b2) show the SJI images covering the evolving ribbon and the nearby sunspot. Same FOV in AIA 171~\AA\ can be found in Fig.~\ref{fig1}~(c2) and (c3), where the ribbon appears fragmented into multiple bright substructures and is superimposed by a funnel-like sunspot plume in the foreground when viewed from above.

Images from Earth view are also compared with \hrieuv high-resolution images from Solar Orbiter's perspective. The \hrieuv pixel scale is 0.49\arcsec\,pixel$^{-1}$. In this dataset, Solar Orbiter has a heliocentric distance of 0.29\,AU, resulting in high spatial resolution of about 200\,km. As shown in Fig.~\ref{fig1}~(d1), \hrieuv FOV covers part of the west limb of the Sun. In this dataset, Solar Orbiter viewed from near the ecliptic plane, with a heliographic latitude of 2.8\textdegree. However, due to the separation angle of about 100\textdegree, the targeted ribbon, especially the western part, is seen very close to the limb. The region covered by SJI is outlined (white) in Fig.~\ref{fig1}~(d1). The western part of the ribbon sub-structure that is covered by IRIS rasters cannot be one-to-one related to the \hrieuv as it is overlapped by both the flare-related filaments in the foreground and the bright plumes in the background. However, to confirm the association between this remote ribbon and the main flare, light curves were computed over selected regions in the SJI and \hrieuv images, indicated by the boxes in Fig.~\ref{fig1}~(b1) and (d2), respectively. The resulting light curves, shown in Fig.~\ref{fig1}~(a2), demonstrate a close temporal correspondence between the ribbon emission and the evolution of the main flare, particularly around the impulsive phase. Although the SJI light curve exhibits multiple peaks and strong fluctuations, likely due to the evolution of individual small-scale brightenings within the relatively small selected region, its intensity enhancements remain closely synchronized with the derivative of GOES flux.

\cite{2026A&A...705A.113C} also reconstruct the HXR sources from simultaneous observations from the Spectrometer/Telescope for Imaging X-rays \citep[STIX;][]{2020A&A...642A..15K} in several energy bands, and identify their spatial relationship with the flare structures. The HXR emission indicates the presence of non-thermal electrons in the flare. However, it is difficult to determine whether a distinct HXR source is associated with our targeted remote ribbon due to the high noise level of the STIX data at this location.

\section{Small-scale downflows}
\label{downflow}

The downflowing structures propagating toward the ribbon locations are well captured in the \hrieuv observations owing to their high temporal and spatial resolution. These features can be seen in greater detail in Fig.~\ref{fig1}~(d2) as slightly elongated bright blobs. \cite{2026A&A...705A.113C} identify these structures as small raining plasma blobs with their spatial sizes at the resolution limit of \hrieuv and short lifetime of 10--20\,s. These bright blobs are separated from each other by only a few 100\,km. Similarly, \citet{2026A&A...711A.279S} study the same ribbon features (termed riblets) in this flare. They report that these riblets typically exhibit velocities of about 100 km\,s$^{-1}$, with lifetimes of about 15\,s and lengths of about 1.6\,Mm.

\begin{figure}[htbp]
\centering 
\includegraphics[width=0.48\textwidth]{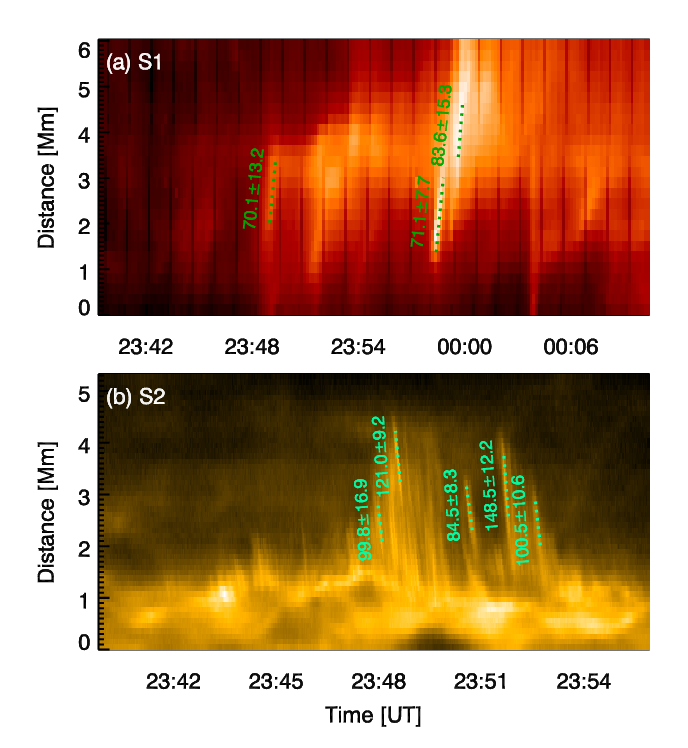}
\caption{Imaging signatures of downflows in the flare ribbon. Time--slice plots generated along slits S1 (a) and S2 (b), as marked in Fig.~\ref{fig1}. Dashed lines highlight examples of flows propagating toward the flare ribbon, with their corresponding POS velocities (in km\,s$^{-1}$) labeled.}
\label{fig2}
\end{figure}

In Fig.~\ref{fig1}~(d3), one of the most elongated downflowing structures (in this flare ribbon) is marked with a blue slit S2. The time-slice plot of this slit can be seen in Fig.~\ref{fig2}~(b), which clearly illustrates the temporal evolution of the downflows. Before the downflows appear, the corresponding footpoint locations are not clearly brightened. The bright flows then appear to move downward, with multiple downflows occurring successively at the same locations. Several of these features are highlighted by dashed lines, with their apparent projected velocities labeled in units of km\,s$^{-1}$. The measured speeds range from 85 to 130\,km\,s$^{-1}$, which indicates rapid apparent motions and is consistent with results in \cite{2026A&A...705A.113C} and \citet{2026A&A...711A.279S}. The velocity and its uncertainty are calculated by identifying the time of maximum intensity at different positions along each marked streak and performing a linear fit of time as a function of distance along the slit. In addition, most of the downflows appear at the lower part of the slit, close to the ribbon location, while the highest ones can extend from over 3\,Mm.

These moving structures propagating toward the ribbon location are also clearly observed in AIA and SJI images, as elongated blobs located south-east of the sunspot (from Earth-view). In Fig.~\ref{fig1}~(b2), one of them is marked with a green slit S1, with the corresponding time-slice plot shown in Fig.~\ref{fig2}~(a). The apparent banding in Fig.~\ref{fig2}~(a) results from the dark vertical line in SJI images corresponding to the location of the spectrograph slit. The blobs can now be seen flowing to the sunspot-ward edges of the ribbon features, with speeds that range from 70 to 85\,km\,s$^{-1}$ in the SJI observations. Note that the downflows identified in the SJI and \hrieuv data are not measured at the same locations, since the SJI FOV only covers the western part of the ribbon, which is masked by absorption features in the \hrieuv observations. In addition, the SJI 1330~\AA\ channel samples plasma at substantially lower temperatures. Nevertheless, the overall temporal coincidence and measured velocities from the two instruments remain consistent.

\section{Spectral diagnostics}
\label{Spectral}

\subsection{Si~{\sc iv}}

The strong Si~{\sc iv} 1393.75 \AA\ and Si~{\sc iv} 1402.77 \AA\ lines forming the Si~{\sc iv} resonance doublet are routinely observed by IRIS \citep{2014SoPh..289.2733D}. In this dataset, the acquired IRIS spectral window only includes the Si~{\sc iv} 1402.77 \AA\ line, thus hereafter simply called Si~{\sc iv} line. The Si~{\sc iv} line generally exhibits a single-Gaussian profile as it is typically formed under optically thin conditions in the transition region, where its emission arises from a relatively narrow temperature range around 8$\times$10$^{4}$\,K \citep{1997ApJ...477L.119D}. In such conditions, the line profile is primarily shaped by thermal broadening and a distribution of unresolved motions (with a velocity component along the line-of-sight (LOS)). When these conditions change, e.g., in flare ribbons, this profile can be modified by additional components, such as blue- or red-wing enhancements associated with high-speed plasma flows or central reversals caused by absorption effects \citep{2015ApJ...811..139T, 2019ApJ...871...23K, 2023ApJS..268...62W}.

\begin{figure*}[htbp]
\centering 
\includegraphics[width=1.0\textwidth]{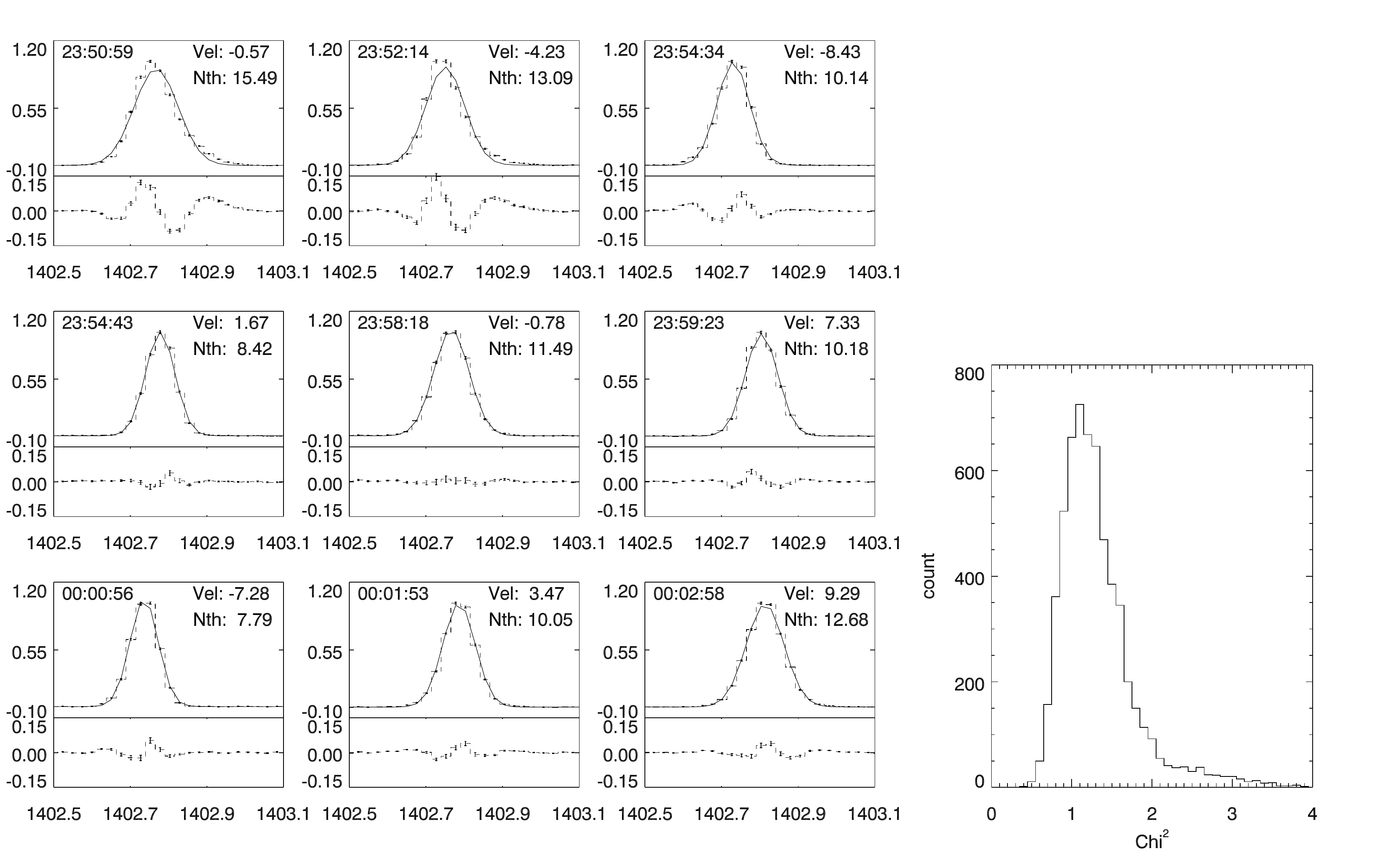}
\caption{Transition region spectral diagnostics (single-Gaussian fitting) of flare ribbons. Normalized Si~{\sc iv} profiles at 9 bright ribbon pixels, along with the single-Gaussian fitting and residuals. The velocity and non-thermal width (in km\,s$^{-1}$) of each profile are labeled in the right upper corner. The lower right panel shows the histogram of the $\chi^2$ values from the single-Gaussian fits for all pixels at the ribbon location.}
\label{fig3}
\end{figure*}

\begin{figure*}[htbp]
\sidecaption
\includegraphics[width=12cm]{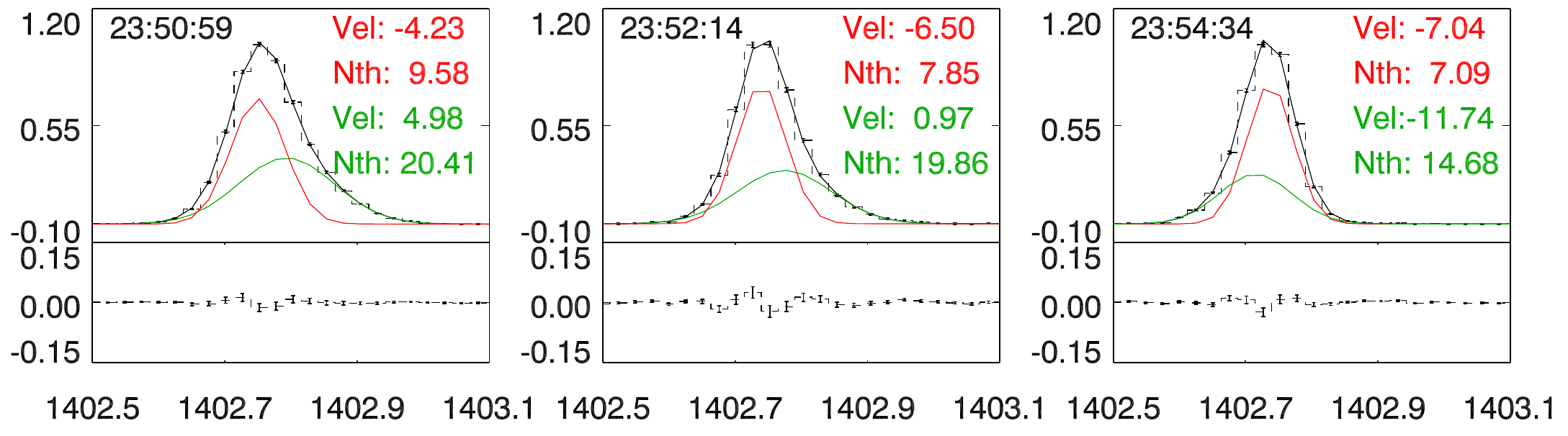}
\caption{Transition region spectral diagnostics (double-Gaussian fitting) of flare ribbons. Normalized Si~{\sc iv} profiles at 3 bright ribbon pixels, same as the pixels shown in the top row of Fig.~\ref{fig3}, along with the double-Gaussian fitting and residuals.}
\label{fig3_duo}
\end{figure*}

In this event, the Si~{\sc iv} profiles at ribbon locations are still generally well described by a single Gaussian. In Fig.~\ref{fig3}, we show the distribution of the $\chi^2$ values from the single-Gaussian fits. The distribution is strongly peaked around 1, suggesting that a single Gaussian provides an adequate description of most of the observed profiles. We also present examples from nine pixels with enhanced brightness compared to their surroundings. Their Doppler velocities ($v_{\rm D}$) are calculated as
\begin{equation}
v_{\rm D} = \frac{\lambda-\lambda_c}{\lambda_c}c,
\end{equation}
where $\lambda_c$ is the resting wavelength 1402.77~\AA\ and c is the light speed. The non-thermal velocity is also presented in km\,s$^{-1}$, as 
\begin{equation}
n_{\rm th} = \sqrt{w_{1/e}^2-w_{th}^2-w_{inst}^2},
\end{equation}
where $w_{1/e}$ is the $1/e$ width ($\sqrt{2}\sigma$) from the Gaussian fitting results. $w_{th}$ is the thermal broadening of Si~{\sc iv} of 6.88\,km\,s$^{-1}$ at its formation temperature of 0.8\,MK and $w_{inst}$ is the instrument broadening ($1/e$ width) of 3.85\,km\,s$^{-1}$ \citep{2014SoPh..289.2733D}. Note that, the calculation of the thermal broadening involves the assumption of isothermal Maxwellian plasma distribution under collisional equilibrium. 

The profiles in the upper row of Fig.~\ref{fig3} show more prominent residuals, indicating that a double-Gaussian fit may provide a better description for some of them. We therefore also apply double-Gaussian fitting to these profiles, with the results shown in Fig.~\ref{fig3_duo}. For these profiles, the double-Gaussian fits are characterized by a strong, narrow component (red) and a weak, broad component (green), which may indicate the presence of more than one velocity component. However, the double-Gaussian fitting does not improve the fit for most pixels, including the other strong profiles shown in Fig.~\ref{fig3}.

Nevertheless, it is clear that these bright ribbon pixels exhibit Si~{\sc iv} profiles that are unexpectedly narrow, with non-thermal widths on the order of 10\,km\,s$^{-1}$ and only show minimal Doppler shifts. These non-thermal widths are a factor 2 smaller than the typical values seen in both quiet sun and active regions \citep[][]{2015ApJ...799L..12D}.

\begin{figure*}[htbp]
\centering 
\includegraphics[width=1.0\textwidth]{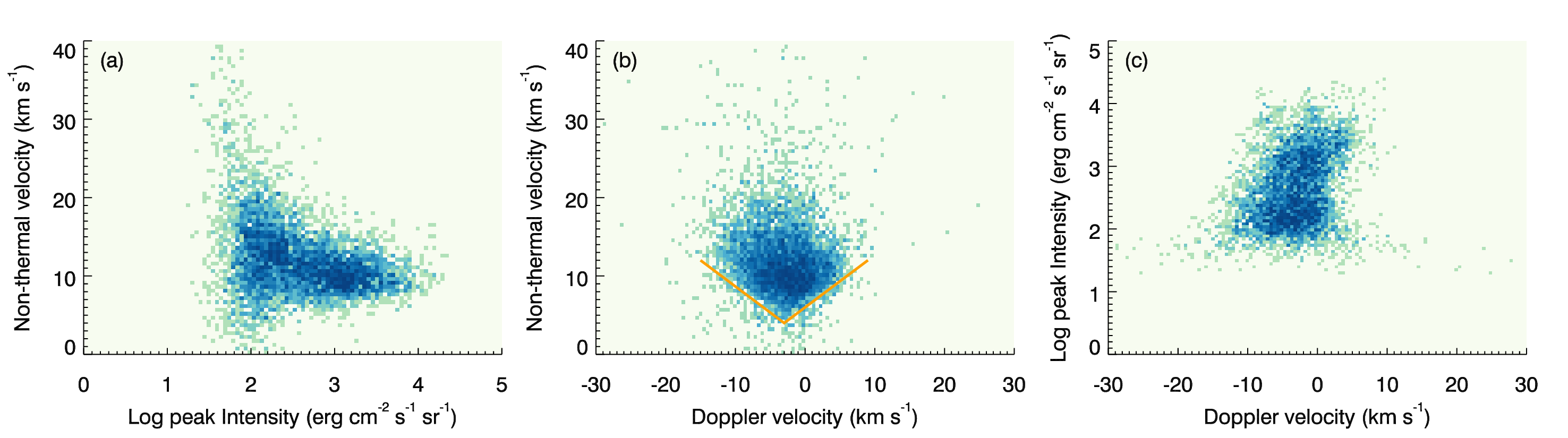}
\caption{Two-dimensional histograms showing the distributions of (a) peak intensity and non-thermal velocity, (b) Doppler velocity and non-thermal velocity and (c) Doppler velocity and peak intensity. The histogram bin sizes are 0.05 for Log peak intensity (erg\,cm$^{-2}$\,s$^{-1}$\,sr$^{-1}$) and 0.5 for both Doppler velocity and non-thermal velocity (km\,s$^{-1}$). The V-shaped orange lines in (b) mark the lower boundary of the distribution, indicating that pixels with larger Doppler velocities rarely exhibit small non-thermal widths.}
\label{fig4}
\end{figure*}

Fig.~\ref{fig4} shows the distribution of peak intensity, non-thermal velocity and Doppler velocity as a function of each other. These distributions are calculated over all pixels in the lower part (from -362.5\arcsec to -339.2\arcsec in solar Y) of the first (from east) four slit positions (from -326.8\arcsec to -320.8\arcsec in solar X), from 23:48 to 00:14~UT. Among these pixels, enhanced brightness is mainly caused by the flare ribbon. 

Panel (a) shows that, at bright ribbon pixels, the non-thermal width becomes less scattered but remains relatively narrow at approximately 10\,km\,s$^{-1}$. A similar behavior has been reported as narrow-line-width UV bursts above sunspots in \cite{2016ApJ...829L..30H}. In contrast, at pixels with lower intensity, the non-thermal width spans a much broader range, which is likely caused by the lower signal-to-noise ratio, introducing greater uncertainty in the fitting process. For the majority of pixels, there appears to be a weak negative correlation between non-thermal width and intensity, although the trend is ambiguous. Nevertheless, even in the absence of a clear correlation, the observed distribution differs from the results obtained in active regions, coronal holes, and the quiet sun \citep{2015ApJ...799L..12D} or microflares \citep[][]{2020ApJ...890L...2C}, which generally show a positive correlation between intensity and non-thermal widths.

Panels (b) and (c) provide additional information on the Doppler velocity distribution. Overall, the Doppler velocities are not particularly pronounced, remaining mostly within 10\,km\,s$^{-1}$. Most pixels, especially those with lower intensities, exhibit blueshifts. At higher intensities, both blueshifted and redshifted components are present. In addition, although no clear correlation is found between non-thermal width and Doppler velocity, the distribution exhibits a V-shaped lower envelope (indicated by the orange lines in panel (b)), such that the minimum non-thermal width increases with the magnitude of the Doppler velocity.

\subsection{C~{\sc ii}}

\begin{figure*}[htbp]
\sidecaption
\includegraphics[width=12cm]{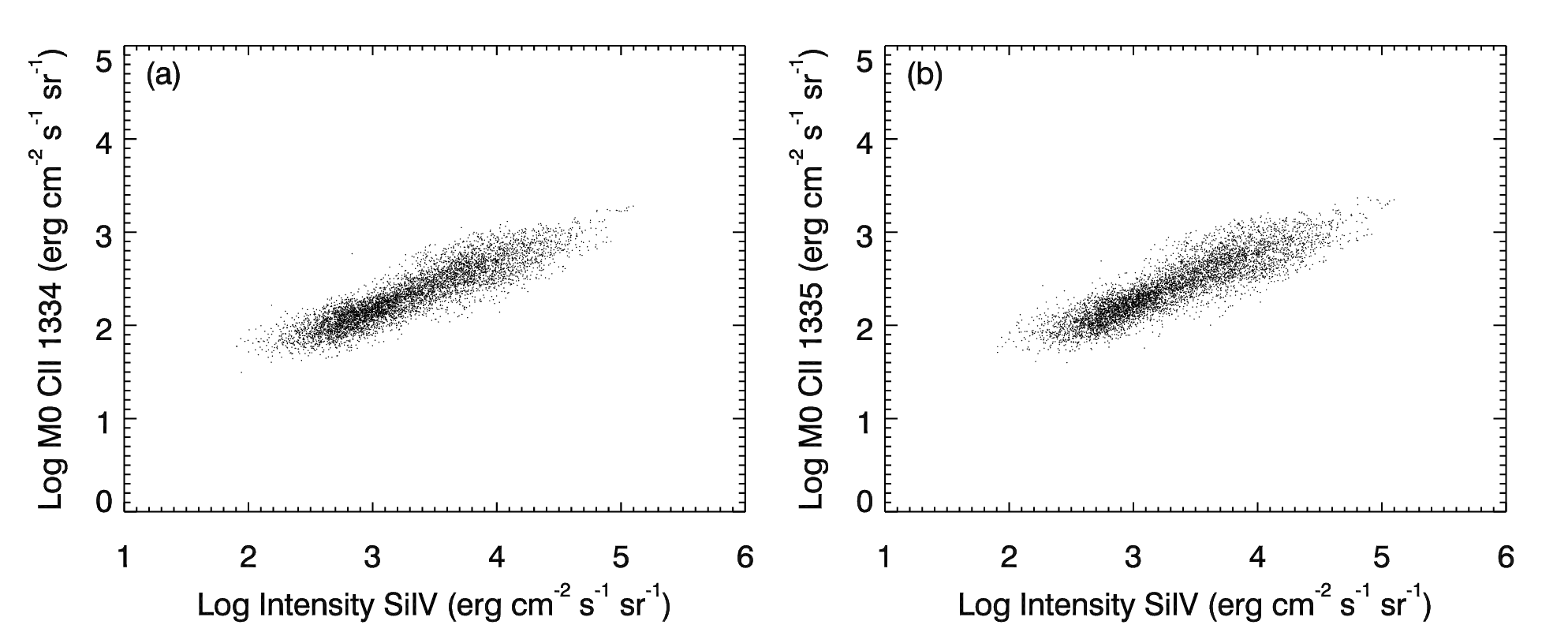}
\normalfont
\caption{Scatter plots showing the correlations between the $M_0$ of the C~{\sc ii} lines and the Si~{\sc iv} intensity derived from Gaussian fitting (integration of the Gaussian profile). Panels (a) and (b) correspond to the C~{\sc ii} 1334.53 \AA\ and 1335.71 \AA\, respectively.}
\label{fig5}
\end{figure*}

C~{\sc ii} 1334.53 \AA\ and 1335.71 \AA\ lines are typically formed at the base or slightly below the transition region at temperatures around 3--5$\times$10$^{4}$\,K. Their emission is almost always optically thick, which leads to their profiles deviating from a simple Gaussian shape, showing instead a flatter top or central absorption. Therefore, we calculate the intensities of the C~{\sc ii} lines by integrating the line profiles after subtracting a constant background derived from nearby continuum wavelengths. This is also the 0th order of moments, written as
\begin{equation}
M_0 = \int I(\lambda)\, d\lambda.
\end{equation}
Fig.~\ref{fig5} shows the correlations between the $M_0$ of C~{\sc ii} lines and the intensity of Si~{\sc iv} as derived from Gaussian fitting. Both panels show clear positive correlations. Similar features can be found in C~{\sc ii}~1335.71~\AA\ and Si~{\sc iv} intensity maps \citep{2015ApJ...814...70R}, which is also shown in our data.

\subsection{Mg~{\sc ii}}

The Mg~{\sc ii} k and h lines sample cooler plasma in the middle to upper chromosphere. Their profiles commonly exhibit double-peaked emission with a central reversal caused by the variation of the source function with height. We apply the {\texttt{iris\_get\_mg\_features}} procedure included in the {\texttt{IRIS sswidl}} package to find the peaks ($k_{\rm 2v}, k_{\rm 2r}, h_{\rm 2v}, h_{\rm 2r}$) and dips ($k_3$, $h_3$). The reversal ratio $R$ is then defined as
\begin{equation}
R = \frac{I_{2v}+I_{2r}}{2I_3} .
\end{equation}

\begin{figure*}[htbp]
\centering 
\includegraphics[width=1.0\textwidth]{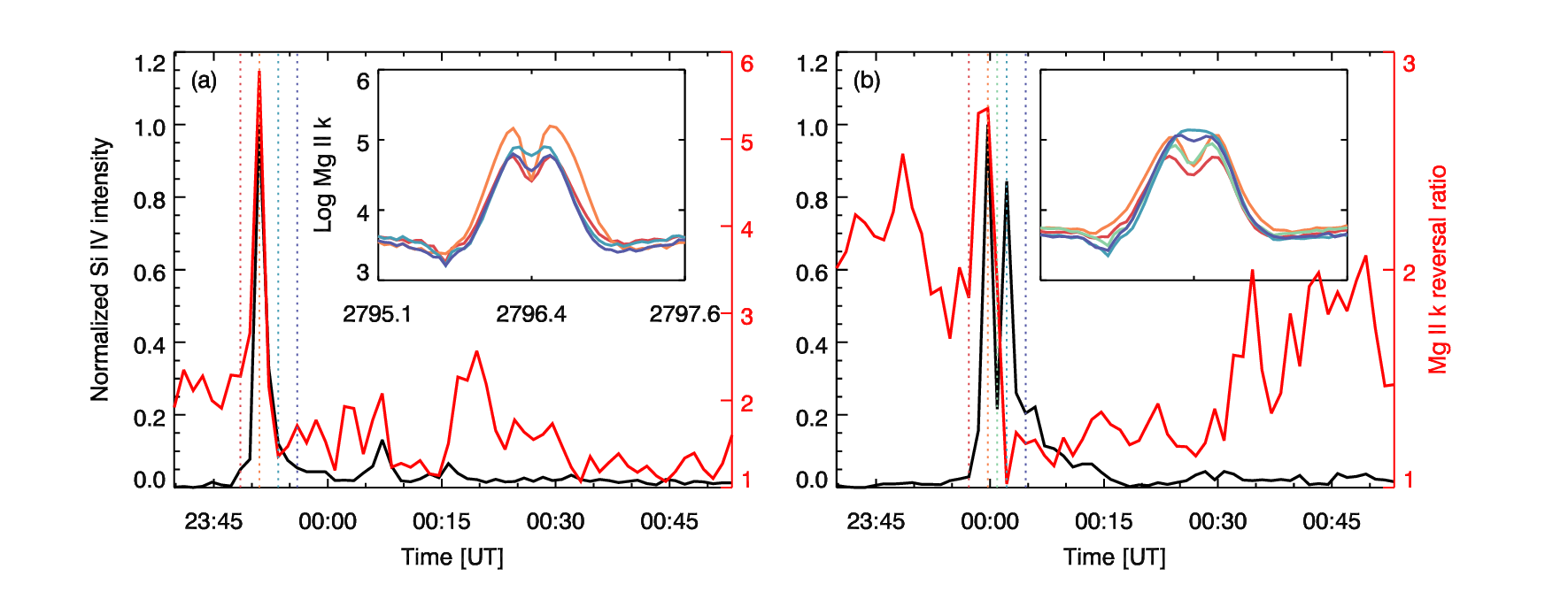}
\caption{Examples of the normalized Si~{\sc iv} intensity (black) and Mg~{\sc ii} reversal ratio (red) as a function of time at bright ribbon pixels. The inset in the upper-right corner shows the Mg~{\sc ii} k line profiles at selected times, which are marked with dotted lines in corresponding colors. In panel (a), the orange profile corresponds to the time of peak Si~{\sc iv} intensity. The pixel shown in panel (b) exhibits two peaks in the Si~{\sc iv} light curve, and the orange and light-blue profiles are sampled at the times of these two peaks, respectively.}
\label{fig6}
\end{figure*}

Fig.~\ref{fig6} shows the normalized Si~{\sc iv} light curve together with the temporal evolution of the Mg~{\sc ii}~k reversal ratio at two ribbon pixels as examples. In panel (a), a clear temporal correlation is observed, with the Mg~{\sc ii} line exhibiting its deepest central reversal at the time of peak Si~{\sc iv} intensity. The Mg~{\sc ii} profiles indicate that this behavior is primarily caused by an enhancement of the $k_{\rm 2v}$ and $k_{\rm 2r}$ peaks during the Si~{\sc iv} peak, while the line core ($k_{\rm 3}$) brightens subsequently, leading to a decrease in the reversal ratio. A similar behavior can also be found in panel (b) as the first peak (orange dashed line) of the Si~{\sc iv} intensity correlates well with the Mg~{\sc ii} reversal ratio. Differently, this pixel brightens up again after 150\,s (light blue dashed line), while the Mg~{\sc ii}~k shows a core ($k_{\rm 3}$) enhancement, which is sufficiently strong that the Mg~{\sc ii} profile transitions into pure emission without a central reversal. This requires sufficient heating in chromospheric plasma, high electron density or strong flows at the core formation height \citep{2017ApJ...842...82R}. 

\subsection{Density diagnostics with O~{\sc iv}}

\begin{table*}[htbp]
\centering
\renewcommand{\arraystretch}{1.2}
\caption{Density estimates over flare ribbon pixels.}
\label{tab1_density}
\begin{tabular}{cccccc}
\hline
No. & Time & Pixel position & O IV ratio & Density ($10^{11}$ cm$^{-3}$) \\
\hline
1 & 23:50:59 & (-320.83\arcsec, -355.39\arcsec) & 0.326 $\pm$ 0.017 & 0.96 -- 1.73\\
2 & 23:54:34 & (-322.87\arcsec, -347.74\arcsec) & 0.385 $\pm$ 0.047 & 1.61 -- 19.00\\
3 & 23:59:23 & (-324.86\arcsec, -347.07\arcsec) & 0.334 $\pm$ 0.041 & 0.76 -- 3.37\\
4 & 00:01:53 & (-324.86\arcsec, -346.74\arcsec) & 0.317 $\pm$ 0.016 & 0.86 -- 1.48\\
5 & 00:02:58 & (-326.87\arcsec, -345.41\arcsec) & 0.361 $\pm$ 0.021 & 1.66 -- 4.30\\
\hline
\end{tabular}
\end{table*}

The Si~{\sc iv} spectral window also contains three of the lines of the O~{\sc iv} 1400~\AA\ multiplet, namely the 1399.77~\AA, 1401.16~\AA\ and 1404.81~\AA\ lines. As they arise from metastable levels, the intensity ratios of these lines can be used as diagnostics of the electron density, especially in the density range $10^{10}$--$10^{12}$~cm$^{-3}$, while being roughly independent from the electron temperature. The O~{\sc iv}~1404.81~\AA\ line is strongly blended with S~{\sc iv}~1404.77~\AA~\citep[e.g.,][]{2001SoPh..200...91T, 2018LRSP...15....5D}. 
Although the contribution from S~{\sc iv}~1404.77~\AA\ can be subtracted from the total intensity using the S~{\sc iv}~1406.08~\AA\ line, which is also covered in this spectral window, it requires an assumption of the intensity ratio between S~{\sc iv}~1404.77~\AA\ and 1406.08~\AA, which is also dependent on the electron temperature. Therefore, we regard the density diagnostics from O~{\sc iv}~1399~\AA\ and 1401~\AA\ to be more reliable. The CHIANTI atomic database version 11.0.2 \citep{1997A&AS..125..149D,2024ApJ...974...71D} is used for the atomic data and density calculations.

For relatively fainter pixels, the signal-to-noise ratio is insufficient to provide good density estimates. We therefore apply the analysis only to 4-pixel binned (along the slit) spectra from the brightest ribbon locations. Table~\ref{tab1_density} summarizes the properties of these pixels, including the O~{\sc iv} intensity ratios $\text{I}_{\text{OIV}}$(1399~\AA)/$\text{I}_{\text{OIV}}$(1401~\AA) and the corresponding electron densities derived under the assumption of an isothermal plasma at a temperature of $10^{5.1}$~K. Note that, for pixel 2, the upper bound of the inferred density range lies beyond the regime where the O~{\sc iv} line ratio remains sensitive to electron density, that is, even a small increase in the measured line ratio can lead to a much larger derived density. Except for this pixel, the derived density ranges from 7$\times$10$^{10}$ to 5$\times$10$^{11}$\,cm$^{-3}$. Compared with typical quiet sun transition region electron densities of 10$^{9}$--10$^{10}$\,cm$^{-3}$, the brightest ribbon locations exhibit density enhancements of approximately one to two orders of magnitude, which is consistent with the intensity enhancement of Si~{\sc iv}. However, this enhancement is modest relative to the main ribbons of strong flares, where densities can reach 10$^{12}$--10$^{13}$ cm$^{-3}$ \citep{2018ApJ...857....5Y}.

\section{Discussions}
\label{Discussions}

\begin{figure*}[htbp]
\centering 
\includegraphics[width=1.0\textwidth]{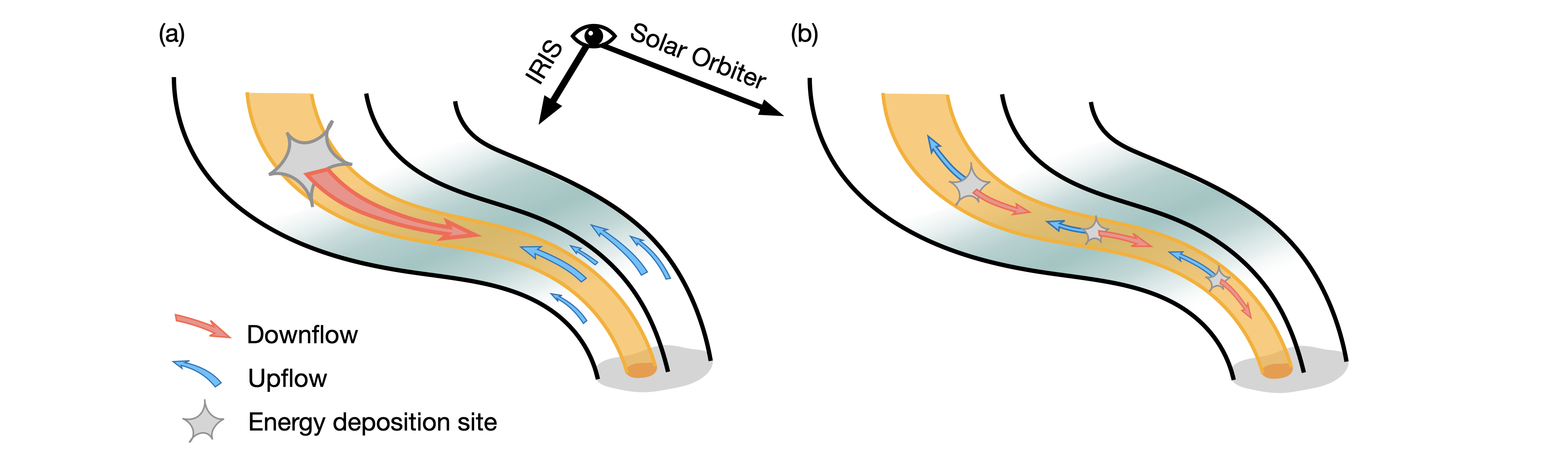}
\caption{Schematic representation of two scenarios. (a) Chromospheric condensation at a greater height produces a downflow, together with a weak but broad background upflowing component. (b) Gradual energy deposition along inclined fields. The shaded regions indicate where the upper chromosphere to mid transition region emission originates, where the magnetic field is predominantly horizontal.}
\label{fig7}
\end{figure*}

Our main results on the flare ribbon downflows can be summarized as follows: (1) the apparent speeds of the small-scale downflowing structures can be measured in both \hrieuv (limb view; 85--130\,km\,s$^{-1}$) and SJI 1330\,\AA\ (disk view; 70--85\,km\,s$^{-1}$) images, while the Si~{\sc iv} line profiles show only very small Doppler shifts; (2) Si~{\sc iv} profiles are systematically narrower in regions of enhanced intensity; (3) Mg~{\sc ii} lines exhibit deeper central reversal at higher Si~{\sc iv} intensities due to enhanced $k_{\rm 2v}$ and $k_{\rm 2r}$ peaks, along with transitions from central reversals to core emission at some locations; and (4) the brightest ribbon pixels show electron densities one to two orders of magnitude higher than typical quiet sun densities.

These results point to a coherent picture of plasma activity in flare ribbons rooted in a strongly inclined magnetic geometry, as the ribbon pixels analyzed here are located close to the sunspot penumbra. The strong magnetic field associated with the sunspot likely expands with height to form a canopy-like structure, guiding plasma motions along highly inclined field lines. This geometry naturally explains the absence of strong Doppler shifts of the Si~{\sc iv} line, despite the clear POS motions observed from two vantage points by Solar Orbiter and IRIS, essentially revealing multi-thermal plasma dynamics over a range of temperatures. Along the inclined chromospheric fields, the flow direction would be nearly perpendicular to the LOS from the IRIS viewpoint, resulting in only a small Doppler component. Moreover, because the field lines are not expected to be exactly perpendicular to the LOS, flows along them can have either a small component toward or away from the observer, depending on the local field orientation. More particularly, since the ribbon is located on the southern side of a sunspot in the southern hemisphere, the connecting field lines can be oriented at an angle slightly greater than 90\textdegree\ to the LOS. Thus, both small blueshifts and redshifts can arise from downflows along the magnetic field, consistent with the Doppler velocity distribution shown in Fig.~\ref{fig4}. Meanwhile, Solar Orbiter observes plasma at a slightly higher altitude, where the magnetic field is expected to become more perpendicular to the solar surface again. This geometry can therefore account for the clear POS motions observed by Solar Orbiter while producing only small LOS velocities in the Si~{\sc iv} emission observed by IRIS.

The same geometry may also explain the unusually narrow line widths. As described in Section~\ref{introduction}, flare ribbons are commonly associated with broadened transition-region line profiles, generally attributed to a variety of effects or physical processes including the superposition of unresolved flows, turbulence, or wave motions. One possible explanation for the narrow profiles observed here is that they primarily originate from a downflow with a narrow LOS velocity distribution, produced by chromospheric condensation following energy deposition at a higher altitude. The narrow width could imply that either the flow itself spans only a limited range of velocities, or that a broader velocity distribution is projected onto a narrower LOS range when the flow is oriented predominantly perpendicular to the LOS. In contrast, non-thermal broadening produced primarily by turbulence would be expected to be less sensitive to projection effects, while transverse wave motions would project more efficiently onto the LOS when the magnetic field is nearly perpendicular to it.

Although not conclusively demonstrated here, the tendency for profiles to become narrower at higher intensities further supports a scenario involving multiple velocity components along the LOS. In addition to the downflow described above, a broader but fainter background component may be present. This component could arise from modest chromospheric evaporation following weak energy deposition at a lower height, or from other chromospheric upflows such as spicules. Because the emission is dominated by the brighter downflow component, the resulting profile remains narrow, and may even be narrower than profiles dominated by the background component. This possibility is also supported by the two velocity components found in the ribbon pixels shown in Fig.~\ref{fig3_duo}. This scenario is illustrated in Fig.~\ref{fig7} (a).

Recent studies have highlighted the importance of magnetic field geometry in determining the observed line broadening and understanding the underlying mechanisms. For example, \cite{2022MNRAS.511.1383R} find that, in the quiet sun, the peak of the Si~{\sc iv} non-thermal velocity distribution increases from approximately 15\,km\,s$^{-1}$ at disk center to 20\,km\,s$^{-1}$ near the limb, consistent with the scenario of unresolved transverse motions across the magnetic field lines which are nearly radial to the surface. In active regions, however, different center-to-limb behavior has been reported. \cite{2024ApJ...975...33C} find that the non-thermal width of Si~{\sc iv} correlates with both redshift and the alignment between the LOS and the magnetic field, leading them to propose that velocity gradients within downflowing plasma are a primary driver of the observed broadening. Similarly, in the coronal temperature regime, \citet{2026arXiv260606577T} conduct a statistical study on flare footpoints and report center-to-limb decrease in all coronal emission line widths, suggesting the broadening is predominately caused by field-aligned motions. Our results are more consistent with these findings that the velocity superposition may be the major cause of the line broadening.

The apparent lower envelope in the width--velocity distribution seen in Figure~\ref{fig4} (b) further supports this picture. Within the magnetic canopy, field lines with different inclinations produce different projected flow velocities. Therefore, flows associated with field lines more closely aligned with the LOS would produce both larger Doppler shifts and a wider LOS velocity distribution.

It is also possible that the apparent downflows do not represent real mass flows but rather the propagation of energy deposition, as illustrated in Figure~\ref{fig7} (b). Instead of losing energy all at once after reaching sufficiently dense plasma, the energy is deposited progressively at different locations, causing each location to brighten in succession, thereby producing the appearance of a moving feature. At each location, the plasma is heated locally and expands, although only weakly, generating both upflows and downflows along the inclined magnetic field. Consequently, the LOS component of these motions remains small, resulting in only weak Doppler shifts and line broadening in the spectral profiles. Given that the energy deposition is localized, the flare ribbon exhibits spatially separated downflows rather than being uniformly brightened. Such a behavior with distinct emission blobs within individual ribbon structures were reported in \cite{2026A&A...705A.113C}.

In addition, the Mg~{\sc ii} profiles indicate that the thermal response of the plasma may be more complex. The evolution of Mg~{\sc ii} profiles shows enhancements of the emission peaks ($k_{\rm 2v}$ and $k_{\rm 2r}$) followed by core emission ($k_{\rm 3}$). A comparable temporal evolution can be found in the 1-D modeling results in \cite{2019ApJ...879...19Z}, where impulsive heating from an electron beam first drives chromospheric downflows before sufficiently increasing the electron density to form single-peak Mg~{\sc ii} profiles. However, in their models, the overall line intensity, including both the $k_{\rm 2}$ peaks and the $k_{\rm 3}$ dip, increases significantly. Therefore, the central reversal is not necessarily deeper than that in the pre-flare profile. In our case, stronger Mg~{\sc ii} emission is generally associated with stronger Si~{\sc iv} emission. This enhancement in Mg~{\sc ii} is mainly due to the enhanced $k_{\rm 2v}$ and $k_{\rm 2r}$ peaks, while the absorption core $k_{\rm 3}$ remains at a similar level (see Figure~\ref{fig6}~(a)) or is only weakly enhanced (see Figure~\ref{fig6}~(b)), resulting in a much deeper central reversal. This suggests that the atmospheric conditions responsible for the cool chromospheric absorption core persist when the Si~{\sc iv} emission becomes enhanced. More recent results from \citet{2023ApJ...944..104P} and \citet{2024ApJ...970...21K} show a similar increase in the depth of the Mg~{\sc ii} central reversal at the ribbon fronts or leading edges of the ribbons. This has been attributed to relatively modest heating deposited deeper in the chromosphere, accompanied by gentle evaporation.

Alternatively, the persistent $k_{\rm 3}$ absorption may indicate that there are low-laying non-flaring loops coexisting with flaring loops. This is in line with recent high-resolution observations that show the presence of uncombed chromospheric loops and non-flaring fine structures spread among flare ribbons \citep[][]{2026ApJ..1005L..68C}. The subsequent enhancement of the Mg~{\sc ii} core emission further suggests that these structures containing cool chromospheric plasma are heated at a later stage, e.g., via cross-field energy transport, which may be more efficient in the partially ionized chromosphere \citep{2015RSPTA.37340268M}.

\section{Conclusions}
\label{Conclusions}

In this work, we investigate small-scale downflowing structures associated with a flare ribbon using stereoscopic observations mainly from \hrieuv aboard Solar Orbiter and IRIS. Although apparent downflow speeds of 70--130 km s$^{-1}$ are observed in both imaging data, the corresponding Si~{\sc iv} spectra exhibit only small Doppler shifts and narrow line profiles in the brightest ribbon regions. The absence of significant LOS velocities and non-thermal broadening is consistent with plasma motions guided by highly inclined magnetic fields and indicates that the brightenings are not accompanied by strong turbulence or wave-driven motions.

This study provides a unique view of the energy deposition site at a flare ribbon under a potentially canopy-like chromospheric environment. However, the magnetic structure of the lower atmosphere remains uncertain. Future chromospheric magnetic field measurements and improved magnetic field extrapolations will be required to test the proposed canopy-like geometry. Extending this analysis to a larger sample of flare ribbons will also help to determine whether the suppressed non-thermal broadening reported here is a common property or a consequence of the local magnetic environment. Another limitation of this study is the lack of simultaneous spectroscopic diagnostics over a wide range of temperatures. While coronal emission can be observed with imagers such as HRI and AIA, their broad temperature response functions and lack of spectral information make it difficult to clearly determine detailed plasma properties. The forthcoming Extreme Ultraviolet High-Throughput Spectroscopic Telescope (EUVST) on board Solar-C \citep{SolarCpaper} will provide high-resolution spectroscopy across a wide temperature range from the chromosphere to the hot corona, and will therefore be highly suitable for future studies of this topic.

Nevertheless, our observations demonstrate that the magnetic conditions of the lower atmosphere could play an important role in shaping the plasma dynamics and spectroscopic signatures observed at flare ribbons. Our findings suggest that the atmospheric response to flare energy deposition could be more complex than implied by simple flare ribbon models and understanding this coupling between magnetic structure and plasma dynamics is essential for developing a more complete picture of energy release in solar flares.

\begin{acknowledgements}
We thank the referee for very constructive comments. The work of Z.H. is funded by the Federal Ministry for Economic Affairs and Climate Action (BMWK) through the German Space Agency at DLR based on a decision of the German Bundestag (Funding code: 50OU2101).
L.P.C. gratefully acknowledges funding from the European Research Council (ERC) under the European Union's Horizon Europe research and innovation program (grant agreement No. 10103984 -- project ORIGIN). Solar Orbiter is a space mission of international collaboration between ESA and NASA, operated by ESA. We thank the ESA SOC and MOC teams for their support. The EUI instrument was built by CSL, IAS, MPS, MSSL/UCL, PMOD/WRC, ROB, LCF/IO with funding from the Belgian Federal Science Policy Office (BELSPO/PRODEX PEA 4000134088, 4000106864 and 4000112292); the Centre National d’Etudes Spatiales (CNES); the UK Space Agency (UKSA); the Bundesministerium f\"{u}r Wirtschaft und Energie (BMWi) through the Deutsches Zentrum f\"{u}r Luft- und Raumfahrt (DLR); and the Swiss Space Office (SSO). SDO is the first mission to be launched for NASA's Living With a Star (LWS) Program and the data supplied courtesy of the AIA consortia. IRIS is a NASA small explorer mission developed and operated by LMSAL with mission operations executed at NASA Ames Research Center and major contributions to downlink communications funded by ESA and the Norwegian Space Centre.
\end{acknowledgements}

%
   \bibliographystyle{bibtex/aa} 
\end{document}